\documentstyle[12pt]{article}
\newcommand{\be}{\begin{equation}}
\newcommand{\ee}{\end{equation}}
\newcommand{\cao}{\c c\~ao\ }

\newcommand{\bt} { \begin{tabular} }
\newcommand{\et}{ \end{tabular} }
\newcommand{\bc} { \begin{center} }
\newcommand{\ec}{ \end{center} }

\newcommand{\f}{ \frac }

\newcommand{\la}{\label }
\newcommand{\bfi}{\begin{figure} }
\newcommand{\efi}{\end{figure} }

\newcommand{\btb} { \begin{table} }
\newcommand{\etb}{ \end{table} }
\begin{document}
\title{THE CRITICAL BEHAVIOUR OF POTTS MODELS WITH SYMMETRY BREAKING FIELDS}

\author{F. C. Alcaraz \ \ and \ \  J. C. Xavier \\
	 Departamento de F\'\i sica \\
	Universidade Federal de S\~ao Carlos \\
	 13565-905, S\~ao Carlos, SP, Brasil }
\date{PACS numbers : 05.50+q,64.60Cn, 75.10Jn }
\maketitle
\vspace{0.2cm}
\begin{abstract}
The $Q$-state Potts model in two dimensions in the presence of external
magnetic
fields is studied. For general $Q\geq3$ special choices of these magnetic
fields produce
effective models with smaller $Z(Q')$ symmetry $(Q'< Q)$. The phase diagram
of these  models and their critical
behaviour are explored by conventional finite-size
 scaling and conformal invariance. The possibility of multicritical behavior,
for finite values of the symmetry breaking fields, in
the cases where $Q>4$ is also analysed. Our results indicate that for effective
models with $Z(Q')$ symmetry $(Q'\leq4)$ the multicritical point occurs at zero
field. This last result is also corroborated by  Monte Carlo simulations.
\end{abstract}
\section{Introduction}
The ferromagnetic $Q$-state Potts model in two dimensions is among the most
studied models of statistical mechanics (see \cite{wu} for a review). In the
absence of external fields the model has a global $Z(Q)$ invariance \cite{2}
which, for low temperatures, is spontaneously broken giving arise to
phase trasitions of second order for
$Q\leq 4$ and first order for $Q > 4$ \cite{baxter}.
The critical fluctuations for $Q=2,3$ and 4 are governed by conformal
field theories with central charges $c=\f{1}{2},\f{4}{5}$ and 1, respectively,
and the whole operator content of these models with several boundary
conditions is known \cite{rittenberg}.

In this paper we study the critical behaviour of these models on the square
lattice, in the presence of symmetry breaking magnetic fields. In general
these fields will break completely the $Z(Q)$ symmetry, but for some special
choices of these magnetic fields the resulting effective
model will have a residual
symmetry $Z(Q')$, with $Q'<Q$, and a domain wall structure at low temperatures,
 similar to these in the $Q'$-state Potts model.

On general grounds \cite{fl},
we do expect that in order-disorder phase transitions, of
discrete symmetry models, the critical behavior is dictated mainly by the
number of ground-states (zero temperature configurations) and the relative
surface energy of the
infinite domain walls connecting these ground states. This reasoning
induces us to expect, for arbitrary values of the symmetry breaking
fields, producing a $Z(Q')$ model, an effectice model in the same
universality class as the $Q'$-state Potts model.
  However this analysis is not valid in general
since some critical and multicritical models, like the Ising model and the
tricritical Ising model \cite{tri}, although having the same number
of ground-states and domain wall structure at low temperatures,
exhibit distinct
critical properties.

An earlier mean field analysis on these models \cite{sf}
indicate that for $Q\leq 4$
 the $Z(Q')$ $(Q'<Q)$ model, produced by the breaking fields, are in the same
 universality class of the $Q'$-state Potts model, for nonzero values of the
fields, while for $Q> 4$, with $Q'< 4$, there is a multicritical point for
finite values of the critical field, where the phase transition changes from
first to second order.

Our study will be done numerically by using standard finite-size scaling
 \cite{fss} to obtain the phase diagram of the models, and the machinery
arising
from conformal invariance \cite{cardy,affleck} to distinguish the several
possible critical behaviours. In the location of multicritical points for $Q>4$
we also perform Monte Carlo some simulations by calculating the fourth-order
cumulant of the magnetization.
\section{The model}
Defining at each lattice site $\vec r=(i,j)$ of a square lattice an integer
variable $n_{\vec r}=0,1,\dots,Q-1$, the Hamiltonian of the $Q$-state Potts
model with $n_{h}$ ($n_{h}=0,1,\dots,Q-1$) symmetry breaking fields
\{\~{h}$_{m}$\}
($m=0,1,\dots,n_{h}-1$) is given by
\be
H_{Q}(\epsilon,\{ \mbox{\~{h}}_{m} \})=
-\epsilon \sum_{<\vec r,\vec r'>} \delta_{n_{\vec r},n_{\vec r'}} - \sum_{m
=0}^{n_{h}-1} \sum_{\vec r} \mbox{\~{h}}_{m}   \delta_{n_{\vec r},m}.
\la{c1.4}
\ee
where $\epsilon>0$ is the ferromagnetic coupling and the first sum runs
over nearest-neighbour sites. In the absence of external fields
the model  has a $Z(Q)$
symmetry, since the configuration  $\{n_{\vec r}\}$ and $ \{n_{\vec r}+ l, mod.
Q \}$
$(l=1,2,\dots,Q)$ has the same energy. The fields \{\~{h}$_{m}$\},
depending on their relative values, break this symmetry totally or partially.
The interesting cases where the symmetry is partially broken are those where
 \~{h}$_{m}$=\~{h} $>0$ ($m=0,1,\dots,n_{h}-1$)
and the remaining symmetry is
$Z(n_{h})\otimes Z(Q-n_{h})$. This symmetry corresponds to $Z(n_{h})$ rotations
among the variables pointing in the field directions and  $Z(Q-n_{h})$
rotations
among the other variables. At zero temperature
we have $n_{h}$ ground states and we do
expect that the $Z(n_{h})$ symmetry is spontaneously broken.

Rather than working with the above Euclidean version of the model
it is convenient to consider its quantum Hamiltonian version
in order to simplify our numerical analysis.
The row-to-row transfer matrix as well the associated $\tau$-continum
quantum Hamiltonian \cite{hq} can be derived by a standard procedure (see
 \cite{chico} for example).
The associated one-dimensional quantum Hamiltonian in
a $L$-site chain is given by
\be
\hat H=-\sum_{l=1}^{L} \sum_{\alpha=0}^{q-1} \biggl(
        (\hat S_{l}\hat S_{l+1}^{\dagger})^{\alpha} +
        \lambda \hat R_{l}^{\alpha} +
        \sum_{m=0}^{n_{h}-1} h_{m} (\hat S_{l} e^{-\frac{2\pi i}{Q} m}
)^{\alpha}
 \biggr),
\la{c1.h}
\ee
where $\lambda$ plays the role of temperature and the magnetic fields
\{$h_{m}$\}
($m=0,1,\dots,n_{h}-1$) are related with the fields
\{\~{h}$_{m}$\} in (\ref{c1.4}). In (\ref{c1.h}) $\hat S_{l}$ and $\hat R_{l}$
are $L^{Q}\otimes L^{Q}$ matrices satisfying the $Z(Q)$ algebra
$$
\begin{array}{cc}
[\hat R_{l},\hat S_{k}]=(\theta-1)\delta_{k,l}\hat S_{k}\hat R_{l}&
\hspace{.5cm} \hat R_{l}^{q}=\hat S_{l}^{q}=1. \\
\end{array}
$$
where $\theta=\exp(\f{i 2\pi}{Q})$, and in
the basis where $\hat S_{l}$ is diagonal
they are given by
$$
\hat S_{l}=1\otimes 1\otimes \cdots 1\otimes S\otimes 1\cdots \otimes 1,
$$
$$
\hat R_{l}=1\otimes 1\otimes \cdots 1\otimes R\otimes 1\cdots \otimes 1
$$
where the matrices $ S$ and $ R$ are in the $l^{th}$ position in
the product and are given by
$$
\begin{array}{cc}
 S=\sum_{i=0}^{Q-1}\theta^{i}|i><i| \mbox{  ,  }&
 R=\sum_{i=0}^{Q-1}|i><[i+1]_{Q}| ,\\
\end{array}
$$
and the symbol $[x+y]_{Q}$ means the addition $(x+y)$, modulo $Q$.

In the absence of magnetic fields (${h}_{m}=0$; $m=0,1,\dots,n_{h}-1$) the
$Z(Q)$ symmetry of (\ref{c1.4}) is reflected in (\ref{c1.h}) by its
commutation with the $Z(Q)$-charge operator
\be
\begin{array}{cc}
\hat P=\prod_{i=1}^{L}\hat R_{l} \mbox{ , } &\hat P^{Q}=1.\\
\end{array}
\la{3}
\ee
The Hilbert space associated with (\ref{c1.h}) can therefore be separated into
disjoint sectors labelled by the eigenvalues $\exp(\f{i 2\pi q}{Q})$,
$(q=0,1,\dots,Q-1)$ of (\ref{3}). The interesting cases, which we will
concentrate on
in this paper,  are obtained by choosing in (\ref{c1.h}) equal values for the
magnetic fields \cite{12}
\be
\begin{array}{cc}
h_{1}=h_{2}=\dots=h_{n_{h}}=h>0.\\
\end{array}
\la{4}
\ee
In this case the symmetry $Z(n_{h})\otimes Z(Q-n_{h})$ of (\ref{c1.4}) is
reflected by the simultaneous commutation of (\ref{c1.h}) with the "parity"
operators
\be
\begin{array}{cc}
\hat V=\prod_{i=1}^{L}\hat V_{l},&\hat V^{n_{h}}=1,\\
\end{array}
\la{vop}
\ee
\be
\begin{array}{cc}
\hat W=\prod_{i=1}^{L}\hat W_{l},&\hat W^{Q-n_{h}}=1,\\
\end{array}
\la{wop}
\ee
with
$$
\hat V_l=1\otimes 1\otimes \cdots 1\otimes V\otimes 1\cdots \otimes 1
$$
$$
\hat W_l=1\otimes 1\otimes \cdots 1\otimes W\otimes 1\cdots \otimes 1
$$
and $V$ and $W$, located at the $l^{th}$ position in the product, are
$Q \times Q$ matrices given by
$$
 V=\sum_{i=0}^{n_{h}-1}|i><[i+1]_{n_{h}}|+
\sum_{i=n_{h}}^{Q-1}|i><i|
$$
$$
 W=\sum_{i=0}^{n_{h}-1}|i><i|+
\sum_{i=n_{h}}^{Q-1}|i><[i+1]_{Q-n_{h}}|.
$$
The Hilbert space associated to  (\ref{c1.h}) is now separated
into $n_{h}(Q-n_{h})$ disjoint sectors
labelled by the eigenvalues $\exp(\f{i 2\pi v}{n_{h}})$
 and $\exp(\f{i 2\pi w}{Q-n_{h}})$
 ($v=0,1,\dots,n_{h}-1$,
   $w=0,1,\dots,Q-n_{h}-1$ ) of the operators $\hat V$ and  $\hat W$,
respectively.

In the numerical diagonalization of (\ref{c1.h}), with periodic boundaries, all
 the above symmetries, together with the
  translational invariance, enables us to handle large lattices with
 modest computer time and memory. We use the Lanczos method to diagonalize
(\ref{c1.h}) up to $L=10,11$ and $13$, for $Q=5,4$ and $3$, respectively.
\section{Results}
We considered in our study only the interesting cases where $Q>n_{h}\geq 2$ and
$h_{1}=h_{2}=\dots=h_{n_{h}}=h\geq 0$, since in these cases always a remaining
 symmetry $Z(Q')$ ($Q'=n_{h}\geq 2$) still remains for $h\neq 0$.

When $h=0$ the model is self-dual with a phase transition at
$\lambda_{c}(0)=1$,
 which has a second order or first order nature
 depending if $Q\leq 4$ or $Q>4$,
repectively. In the limit $h\rightarrow\infty$ the eigenvectors of
 $\hat S_{i}$ in
 (\ref{c1.h}) with eigenvalues $\theta^{l}$, $l=n_{h},n_{h}+1,\dots,Q-1$,  are
forbidden and we have an effective  $n_{h}$-state Potts model at zero
field.
Analysing the effect of (\ref{c1.h}) in the remaining $n_{h}^{L}$-dimensional
Hilbert space it is not difficult to see that the phase transition happens
at
\be
\lambda_{c}(h\rightarrow\infty)=\f{Q}{n_{h}}
\la{pt}
\ee
Between those two extremum values of $h$ we estimate the phase transition curve
$\lambda_{c}(h)$ by using standard finite-size scaling. The curve is evaluated
 by extrapolations to the bulk limit
($L\rightarrow\infty$) of sequences
$\lambda_{c}(h,L)$ obtained by solving \cite{fss}
\be
\begin{array}{cc}
\Gamma_{L}(\lambda_{c})L=\Gamma_{L+1}(\lambda_{c})(L+1),& L=2,3\dots , \\
\end{array}
\la{gap}
\ee
where $\Gamma_{L}(\lambda_{c})$ is the mass gap of the Hamiltonian (\ref{c1.h})
with  $L$ sites.

Once the transition curve is estimated, in the region of continuous phase
transitions  we expect the model is conformally invariant.
This symmetry allows us
to infer the critical properties from the finite-size corrections to the
eigeinspectrum at $\lambda_{c}$ \cite{cardy}. The conformal anomaly $c$ can be
calculated from the large $L$ behaviour of the ground-state energy $E_{0}(L)$.
For periodic chains $E_{0}(L)$ behaves as
\be
\frac{E_{0}(L)}{L}= \epsilon_{\infty} - \frac{\pi c v_{s}}{6L^{2}} +o(L^{-2}),
\la{ein}
\ee
where $\epsilon_{\infty}$ is the ground-state energy, per site, in the
bulk limit and $v_{s}$ is the sound velocity. The scaling dimensions  of
operators governing the critical fluctuations  (related to critical exponents)
are evaluated from the finite-$L$ corrections of the excited states. For each
 primary operator, with dimension $x_{\phi}$ and spin $s_{\phi}$, in the
operator algebra of the system, there exists an infinite tower of eigenstates
of the quantum Hamiltonian, whose energy $E_{m,m'}^{\phi}$ and momentum
 $P_{m,m'}^{\phi}$, in a periodic chain are given by
\be
\begin{array}{l}
E_{m,m'}^{\phi}(L)=E_{0}+\frac{ 2\pi v_s }{ L }(x_{\phi}+m+m') + o(L^{-1}) \\
P_{m,m'}^{\phi}=(s_{\phi}+m-m') \frac{2 \pi}{L}\\
\end{array}
\la{ep}
\ee
where $m,m'=0,1,\dots$ .

We present our results separately for the cases $Q\leq 4$ and $Q>4$ in
the next sections.
\subsection{Models with $Q\leq 4$}
There exist three interesting cases, namely, the 3-state Potts model with
two fields ($Q=3,n_{h}=2$) and the 4-state Potts model with three and two
fields ($Q=4,n_{h}=3,2$).

The critical curves were obtained by solving (\ref{gap}). As an example, in
Fig.1 we show the extrapolated curve for the case of $Q=3$ and $n_{h}=2$. We
 also show in this figure the curve obtained by solving (\ref{gap}) for
$L = 5$. We clearly see  an agreement with the limiting values
$\lambda_{c}(0)=1$ and $\lambda_{c}(h\rightarrow\infty)=\f{3}{2}$, predict by
(\ref{pt}). Similar curves are obtained in the other cases.

The conformal anomaly and anomalous dimensions are obtained using relations
(\ref{ein},\ref{ep}), for several values of $h$. In table 1,2 and 3 we show for
some values of $h$ the extrapolated results obtained for the cases
($Q=3,n_{h}=2$) ($Q=4,n_{h}=3$) and
 ($Q=4,n_{h}=2$), respectively. Our personal estimative of  errors are
in the last digit. In these tables
we also present our conjectured values. The dimensions $x_{n}(k,v)$ and
$x_{n}(k,v,w)$ appearing in these tables are obtained
by using in (\ref{ep}) the $n$th $(n = 1,2,3,\ldots)$ eigenenergy
 in the sector with
momentum $\f{2\pi k}{L}$ ($k=0,1\dots$) and eigenvalues
$\exp(\f{i 2\pi v}{Q-n_{h}})$ and $\exp(\f{i 2\pi w}{Q-n_{h}})
 $($v=0,1,\dots,n_{h}-1$,
 $w=0,1,\dots,Q-n_{h}-1)$ of the operators
$\hat V$ and $\hat W$ defined in (\ref{vop}) and (\ref{wop}), respectively.

We see in table 1 that for all values of $h$ the conformal anomaly is
 $c=\f{1}{2}$, indicating that the model share the same universality class  as
the $Z(2)$ Ising model. The dimensions 1 and $\f{1}{8}$ correspond to the
dimensions of the energy and magnetic operator in the Ising model and the
dimension 2=1+1 and $\f{9}{8}=1+\f{1}{8}$ are the next dimensions in the
tower of these operators (see Eq. (\ref{ep})).

In table 2 we clearly see that the conformal anomaly, for all values of h is
$c=\f{4}{5}$. There exists two modular invariant universality classes of
conformal theories with $c=\f{4}{5}$ \cite{ciz}. One of these can
 be represent by the
restricted solid-on-solid (RSOS) model \cite{rsos1,rsos2} and the other
by the 3-state Potts model with no
magnetic fields. These two models, although having the same conformal
anomaly $c=\f{4}{5}$,
have distinct operator content. The dimension $x=0,\f{4}{5},\f{14}{5},\f{2}{5},
\f{4}{3}$ in table 2 and
the degeneracy of the sectors where the operator $\hat V$ in
 (\ref{vop}) has eigenvalues $\exp(\f {i2\pi}{3})$ and
 $\exp(\f {i4\pi}{3})$
indicate that the model belongs to the same universality class as the
3-state Potts model.

In table 3, as in table 1, the conformal anomaly is $c=1/2$ and the dimensions
are those of the Ising model indicating that both models are in the same
universality class.

Beyond the values of $h$ presented in tables
 1-3 we also performed a carefull analysis
for small values of $(h \sim 0.01)$ finding  similar results as those
presented in those tables. This indicate that for $Q\leq4$, where the model
 has a second-order phase transition in the absence of external
fields, the introduction of $n_{h}$ ($Q>n_{h}\geq2$) fields,
of arbitrary strength,
brings the model into the universality class of a Potts model with
$n_h$ states.
\subsection{Models with $Q>4$}
In this case, while in the absence of external fields the models exhibit a
first-order phase transition the introduction of $n_{h}$ magnetic fields
($ 4\geq n_{h}\geq2$) of infinite and equal strength $(h\rightarrow\infty)$
 render them to an effective $n_{h}$-state Potts model, which has a second
order phase transition. This brings the interesting possibility of a
multicritical behaviour for a finite value of $h$, when the transition curve
changes  from second to first order as we decreases $h$ from the infinite
value. In fact this is the mean field prediction \cite{sf}.

Since the Hilbert space grows exponentially with $Q$, the simplest case where
the above critical point may occur is the 5-state Potts model in the
presence of $n_{h}=2$ magnetic fields. In table 4 we present, for some values
of $h$, our results for the finite-size sequences of the conformal anomaly
of the model. These sequences are obtained from (\ref{ep}) and (\ref{ein})
\be
c_{L,L+1}=
\f{ 12[(L+1)E_{0}(L)-LE_{0}(L+1)] }{ [(L+1)^{2}-L^{2}][E_{2}(L+1)-E_{1}(L+1)] }
\la{c}
\ee
where $E_{0}(L)$ is the ground-state energy for the chain of length $L$
and $E_{2}(L)$, $E_{1}(L)$ are the lowest eigenergies with momentum 0 and
$\f{2\pi}{L}$, respectively, in the sector where the operators $\hat V$ and
$\hat W$, defined in (\ref{vop}) and (\ref{wop}) has eigenvalues (-1) and (1).
We see in table 4 that for
$h$\put(5,2.5){$>$}\put(5,-3.3){$\sim$}\hspace*{.6cm}0.05 we still have
an Ising-like
behaviour with $c=\f{1}{2}$. For $h<0.05$,
 and for the lattice sizes we were
able to handle, it is not possible to obtain reliable results using (\ref{c}).

An heuristic method, which was proved to be effective in obtaining
multicritical
points in earlier works \cite{sucesso} is to simultaneously solve
Eq. (\ref{gap}) for three different lattice sizes
\be
\Gamma_{L}(\lambda_{c})L=\Gamma_{L+1}(\lambda_{c})(L+1)=
\Gamma_{L+2}(\lambda_{c})(L+2) .\\
\la{2gap}
\ee
We tried to solve these equations for $0.5>\lambda>0$ ($L=5$) and we
found no consistent solutions, which indicates the absence of a tricritical
point for a finite value of $h$.

Another method, also used to locate multicritical  points \cite{sucesso}, is
obtained from the simultaneous crossing of two different gaps on a given
pair of lattices (instead of three lattices
as in (\ref{2gap})). Trying several
different gaps we also did not find, within this method, a multicritical
point for $h\neq 0$.

Since these methods are heuristic and the lattice sizes we are considering
may not be enough to obtain the bulk limit ($L\rightarrow\infty$) in the
region $h\sim 0$ we decide to supplement our results by Monte Carlo
simulations.
 These simulations will enable us
 to distinguish the order of the phase transition  as we change the
magnetic field strength. We simulate the systems by the heath-bath
algorithm and analyse the fourth-order cumulant of the
magnetization as a function of the magnetic field. The simulations
were done directly in the
classical version of the model (\ref{c1.4}). Since the evidence of a
multicritical point, for non-zero values of $h$, would be large for
higher values
of $Q$, we choose $Q=7$ and $n_{h}=2$ (\~{h}$_{0}$=\~{h}$_{1}=h$) for extensive
calculations.

The fourth-order cumulant of the magnetization
is defined by
\be
U_{L}=1-\f{<m^{4}>_{L}}{3<m^2>^{2}_{L}}
\la{cumu}
\ee
where the averages are done on an $L \times L$ lattice. The magnetization in
(\ref{cumu}), for a given configuration $\{n_{\vec r}$\}
of classical variables, is
defined by
$$
m=\frac{1}{L^2} \sum_{\vec r} (\delta_{n_{\vec r},0}-\delta_{\vec r,1} )
$$
Following Binder \cite{binder}, the cumulant (\ref{cumu}) will be zero for
$T>T_{c}$ and $U_{L}=\f{2}{3}$ for $T<T_{c}$. At the transition
temperature $T_{c}$ (\ref{cumu}) will be zero for continous phase transitions
 and
negative for first-order phase transitions.

In Fig. 2 we  show the values of $U_{L}$ for lattice sizes $50\times 50,
80\times 80$ and $170\times 170$. In the simulations we choose $\epsilon=1$,
\~{h}=0.01 and each point was obtained by averaging $5.10^{4}$ iterations,
after thermalization. We see in this figure that while for $L=50$ the
phase transition appears to be first order, as $L$ grows the numerical
results indicates the phase transition to be continuous. The result for the
 smaller lattice $L=50$ is clearly due to the finite size of the lattice.
By repeating these simulations for even smaller values of \~{h} we should
expect that these finite-size effects will be apparent for even larger
lattices, and our simulations are in favour of a multicritical point only at
\~{h}=0.
\section{Conclusion}
%
We have calculated the phase transition diagram and critical properties of the
$Q$-state Potts model in the presence of $n_{h}$ $(Q>n_{h}>1)$ external
magnetic fields of equal strengh $h>0$. In the case where $Q>n_{h}\geq 2$
the original symmetry, at $h=0$, breaks into a $Z(n_{h})\otimes Z(Q-n_{h})$
symmetry. The $Z(n_{h})$ part of the above symmetry relates the configurations
of the $n_{h}$ distinct ground-state configurations at zero temperature,
and by standard arguments, should be spontaneously broken at low temperatures.

Our results, based on conformal invariance and supplemented by Monte Carlo
simulation indicate that, for arbitrary values of $h$, the order-disorder
phase transition associated with the global $Z(n_{h})$ symmetry is in the same
universality class of the $n_{h}$-state Potts model. Morever, for $Q>4$,
contrary to the mean field prediction, we do not see any evidence of a
 multicritical point for non-zero values of $h$.
\begin{center}
{\bf Acknowledgments}
\end{center}

We thank M.T.  Batchelor for a careful reading of our manuscript. This
work was support in part by Conselho Nacional de
Desenvolvimento
Cient\'{\i}fico e Tecnol\'ogico - CNPq-Brazil and by Funda\cao de Amparo \`a
Pesquisa do Estado de S\~ao Paulo-FAPESP-Brasil.
\newpage
\Large
Figure Captions
\normalsize
\vspace{1.5cm}

Fig. 1 - Estimates for the critical curve of the 3-state Hamiltonian
(\ref{c1.h}) with $n_{h}=2$ fields. The  curve in the largest scale, for
$0< h < 5$, interpolates the points
obtained by extrapolating the solutions of (\ref{gap}) for $L=2-13$ (circles).
The inserted curve for $0 < h < 50$ was obtained by solving
(\ref{gap}) for $L = 5$.
\vspace{1.0cm}

Fig. 2 - Fourth-order cumulant of the magnetization (\ref{cumu}) as a function
 of $\beta$ for the 7-state Potts model in presence of $n_{h}=2$ external
 fields  \~{h}$_{0}$=\~{h}$_{1}$=0.01 (see Eq. (\ref{c1.4})). The lattice sizes
 are $L=50\times 50$, $L=80\times 80$ and $L=170\times 170$.
\vspace{1.0cm}
\newpage
\Large
Table Caption
\normalsize
\vspace{1.5cm}

Table 1 - Extrapolated and conjectured results for the conformal anomaly $c$
and
 anomalous dimensions $ x_{n}(k,v)$ of the 3-states Potts chain (\ref{c1.h})
with $n_{h}=2$ magnetic fields ($h_{0}=h_{1}=h$) (see the text). The
conjectured
 values, in parenthesis, are the corresponding ones in the critical Ising
model.
\vspace{1.5cm}

Table 2 - Extrapolated and conjectured results for the conformal anomaly $c$
and
 anomalous dimensions $ x_{n}(k,v)$ of the 4-states Potts chain (\ref{c1.h})
with $n_{h}=3$ magnetic fields ($h_{0}=h_{1}=h_{2}=h$) (see the text).
The conjectured
 values, in parenthesis, are the corresponding ones appearing
in the 3-state Potts model.
\vspace{1.5cm}

Table 3 - Extrapolated and conjectured results for the conformal anomaly $c$
and
 anomalous dimensions $ x_{n}(k,v,w)$ of the 4-states Potts chain (\ref{c1.h})
with $n_{h}=2$ magnetic fields ($h_{0}=h_{1}=h$) (see the text).
The conjectured
 values, in parenthesis, are the corresponding ones in the critical Ising
model.
\vspace{1.5cm}

Table 4 - Finite-size sequences $c_{L,L+1}$, defined in (\ref{ein}) for the
5-state Potts model Hamiltonian (\ref{c1.h}) with $n_{h}=2$ ($h_{0}=h_{1}=h$)
external magnetic fields. The last line gives the extrapolated results.
\newpage
\Large
\bc
Table 1
\ec
\normalsize
\btb[h]
\bc
\bt{||c|l|l|l|l|l||} \hline\hline
 h   & c        &$x_{2}(0,0)$ & $x_{1}(1,0)$ &$x_{1}(0,1)$ &$x_{1}(1,1)$\\
\hline
 0.5 &0.50087   &1.00073       &2.001       &0.1251       &1.1251    \\
     & \hspace{.3cm} (0.5)   & \hspace{.5cm}(1) & \hspace{.4cm}(2)   &  (0.125)
   & (1.125)   \\ \hline
 1.0 &0.50000   &0.99996       &1.9995      &0.12497     &1.12497 \\
    &  \hspace{.3cm}(0.5)   &  \hspace{.5cm}(1)& \hspace{.4cm}(2)&  (0.125)
& (1.125)   \\ \hline
 1.5 &0.50003  &1.00000     &1.99997      &0.1249      &1.1249   \\
    & \hspace{.3cm}(0.5)   & \hspace{.5cm}(1)& \hspace{.4cm}(2)&  (0.125)    &
(1.125)   \\ \hline
 2.0 &0.50008   &1.00000     &2.00000       &0.12500      &1.12500 \\
    &  \hspace{.3cm}(0.5)   & \hspace{.5cm}(1) & \hspace{.4cm}(2)&  (0.125)
& (1.125)   \\ \hline
\et
\ec
\etb
\vspace{3.cm}

\Large
\bc
Table 2
\ec
\normalsize
%
\btb[h]
\bc
\bt{||c|c|c|c|c|c|l|l||} \hline
 h   & c        &$x_{2}(0,0)$ & $x_{3}(0,0)$ &$x_{2}(0,1)$ &$x_{3}(0,1)$  &
$x_{1}(1,1)$&$x_{2}(1,1)$ \\ \hline\hline
 0.5 &0.800     &0.8001       &2.79          &0.1333     &1.332          &
1.1333     & 2.332 \\
     &(0.8)     &(0.8)      & (2.8)       &  (0.133...) & (1.33...)      &
(1.133...)   & (2.33...) \\ \hline
 1.0 &0.7991   &0.7997       &2.8            &0.1329      &1.330          &
1.1329      & 2.331 \\
     &(0.8)    &(0.8)     &  (2.8)        &  (0.133...) & (1.33...)      &
(1.133...)   & (2.33...) \\ \hline
 1.5 &0.79     &0.799        &2.83           &0.1334      &1.330          &
1.1334      & 2.3319 \\
    &(0.8)     &(0.8)     &  (2.8)        &  (0.133...) & (1.33...)      &
(1.133...)   & (2.33...)  \\ \hline
 2.0 &0.7991   &0.7999       &2.84           &0.13345      &1.330          &
1.13345    & 2.3319 \\
    &(0.8)     &(0.8)     &  (2)          &  (0.133...) & (1.33...)      &
(1.133...)   & (2.33...)\\ \hline
\et
\ec
\etb
%
%

\newpage
\Large
\bc
Table 3
\ec
\normalsize
%
\btb[h]
\bc
\bt{||c|c|c|c|c|c||} \hline
 h   & c        &$x_{2}(0,0,0)$ & $x_{1}(1,0,0)$ &$x_{1}(0,1,0)$
&$x_{1}(1,1,0)$\\ \hline\hline
 0.5 &0.5002   &0.997       &1.944     &0.1248       &1.1248      \\
     &  (0.5)   &   (1)        &  (2)       &  (0.125)    & (1.125)   \\ \hline
 1.0 &0.505   &1.0000       &1.999      &0.12509     &1.12509    \\
    &  (0.5)   &   (1)        &  (2)       &  (0.125)    & (1.125)   \\ \hline
 1.5 &0.5000   &0.9999      &1.9996       &0.1249      &1.1249      \\
    &  (0.5)   &   (1)        &  (2)       &  (0.125)    & (1.125)   \\ \hline
 2.0 &0.5000   &0.9999      &1.9997       &0.12499     &1.12499 \\
    &  (0.5)   &   (1)        &  (2)       &  (0.125)    & (1.125)   \\ \hline
\et
\ec
\etb
%
%
%
%
%
\vspace{2.5cm}

\Large
\bc
Table 4
\ec
\normalsize
%
%
\btb[h]
\bc
\bt{||c|c|c|c|c||} \hline
 $ _{\makebox[.5cm]{\large $N$}} \setminus^{ \makebox[.5cm]{\large $h$} }$    &
 0.05  &  0.1 & 0.5 & 1.0 \\ \hline\hline
 4   &1.494578  &1.293239  &0.822347  & 0.747102  \\ \hline
 5   &1.150201  &0.947010  &0.665147  & 0.632826    \\ \hline
 6   &0.957555  &0.784083  &0.603697  & 0.583638  \\ \hline
 7   &0.834717  &0.696948  &0.572906  & 0.558058   \\ \hline
 8   &0.753559  &0.646537  &0.554626  & 0.542796    \\ \hline
 9   &0.698775  &0.615352  &0.542598  & 0.532888    \\ \hline
10  &0.660931  &0.594802  &0.534149  &  0.526071   \\ \hline\hline
$\infty$ &0.5693  &0.5448       &0.4898     &0.500\\ \hline
\et
\ec
\etb
%

\begin{thebibliography}{99}
%
%
%
%
%
\bibitem{wu}
 Wu F 1982
{\it Rev. Mod. Phys }
{\bf 54} 235
%
%
%
%
\bibitem{2}
Actually the model has a larger symmetry $S(Q)$, the permutation group of
$Q$ objects.
%
%
%
%
\bibitem{baxter}
Baxter R J (1982) {\it Exactly Solved Models in Statistical Mechanics}
  (New York: Academic)

%
%
\bibitem{rittenberg}
 von Gehlen G, Rittenberg V and Ruegg H (1986)
 {\it J. Phys. A: Math Gen}
{\bf 19} 107 ; see also [8] and [13].
%
%
\bibitem{fl} Fr\"olich J and Lieb E H (1979)
{\it Comm. Math. Phys.} {\bf 60} 2338
%
%
\bibitem{tri}
Lawrie I D and Sarbach S (1984) {\it Phase Transitions and Critical Phenomena}
vol 9, ed C Domb and J L Lebowitz \ \  \ (New York: Academic) p 1\\
Knobler C M and
Scott R L (1984) {\it Phase Transitions and Critical Phenomena} vol 9,
ed C Domb and J L Lebowitz \ \  \ (New York: Academic) p 164
%
%
%
%
\bibitem{sf}
 Straley J P and Fisher M E (1973) {\it J. Phys. A: Math Gen },
{\bf 6} 1310
%
%
\bibitem{fss}
Barber M N (1983) {\it Phase Transitions and Critical
 Phenomena} vol 8, ed C Domb and J L Lebowitz
  (New York: Academic) p 145
%
%
%
%
\bibitem{cardy}
Cardy J L (1987) {\it Phase Transitions and Critical
 Phenomena} vol 11, ed C Domb and J L Lebowitz
  (New York: Academic ) p 55
%
%
\bibitem{affleck}
Bl\"ote H W J, Cardy  J L and Nightingale M P (1986)
{\it Phys. Rev. Lett. }
{\bf 56} 742 \\
Affleck I (1986)
{\it Phys. Rev. Lett. }
{\bf 56} 746
%
%
%
\bibitem{hq}
 Fradkin E and Susskind L (1978)
{\it Phys. Rev. }D
{\bf 17} 2637
%
%
\bibitem{chico}
 Alcaraz F C and  K\"oberle R (1981)
 {\it J. Phys. A: Math Gen }
{\bf 14} 1169
%
%
\bibitem{12}
The cases where $h<0$ are physically equivalent to \ the \ choices
$n_{h}'=Q-n_{h}$,
with $h'>0$.
%
%
\bibitem{ciz}
 Cappelli A, Itzykson C and Zuber J-B (1987)
{\it Nucl. Phys.} B
{\bf 280} 445 ; (1987) {\it Comm. Math. Phys.} {\bf 113} 1
%
%
\bibitem{rsos1}
 Andrews E, Baxter R J and Forrester P J (1984)
{\it J. Stat. Phys }
{\bf 35} 193
%
%
\bibitem{rsos2}
 Huse D A (1984)
{\it Phys. Rev.}  B
{\bf 30} 3908
%
%
\bibitem{sucesso}
 Fisher M E, and Berker N (1982)
{\it Phys. Rev.}
 {\bf 26} 2707 \\
Rikvold P A, Kinzel W, Gunton J D and Kaski K (1983)
{\it Phys. Rev} B {\bf 28} 2686 \\
Hermann H J (1984)
{\it Phys. Lett.} A {\bf 100} 156 \\
Alcaraz F C, de Fel\'{\i}cio J R D, K\"oberle R and Stilck J F  (1985)
{\it Phys. Rev.}  B
 {\bf 32} 11 \\
 Malvezzi A L (1994)
{\it Braz. J. Phys.}
 {\bf 24} 2
%
%
\bibitem{binder}
 Binder K (1981)
{\it  Phys. Rev. Lett.}
 {\bf 47} 693 ; (1981)
 {\it Z. Phys. } B
 {\bf 43} 119 (1981) \\
Binder K and Landau L P (1984) {\it Phys. Rev.} B {\bf 30} 1477
\end{thebibliography}
\end{document}